\documentclass[showpacs]{revtex4}
\usepackage{epsfig}
\usepackage{graphicx}
\usepackage{subfigure}
\usepackage{dcolumn}
\usepackage{amsmath}
\usepackage{latexsym}

\newcommand{\afg}[2]{\frac{d {#1}}{d {#2}}}
\newcommand{\pafg}[2]{\frac{\partial {#1}}{\partial {#2}}}
\newcommand{\ket}[1]{\left|#1\right\rangle} 
\newcommand{\bra}[1]{\left\langle#1\right|}

\newcommand{\braket}[2]{\left\langle#1|#2\right\rangle}
\newcommand{\rbraket}[2]{\left(#1|#2\right)}
\begin{document} 

\begin{flushright}
\end{flushright}

\title{Bound state energies and phase shifts of a non-commutative well}  
\date{\today}
\author{J D Thom$^{b}$\footnote{e-mail: nthom@sun.ac.za} and
F G Scholtz $^{a,b}$\footnote{e-mail:
fgs@sun.ac.za}}
\affiliation{$^a$National Institute for Theoretical Physics (NITheP), 
Stellenbosch 7600, South Africa\\
$^b$Institute of Theoretical Physics, 
University of Stellenbosch, Stellenbosch 7600, South Africa}

\begin{abstract}
Non-commutative quantum mechanics can be viewed as a quantum system represented in the space of Hilbert-Schmidt operators acting on non-commutative configuration space. Within this framework an unambiguous definition can be given for the non-commutative well.  Using this approach we compute the bound state energies, phase shifts and scattering cross sections of the non-commutative well.  As expected the results are very close to the commutative results when the well is large or the non-commutative parameter is small.  However, the convergence is not uniform and phase shifts at certain energies exhibit a much stronger then expected dependence on the non-commutative parameter even at small values. 
\end{abstract}

\maketitle

\section{Introduction}
The idea of non-commutative spacetime was first formally
introduced by Snyder in \cite{snyder} as an attempt to regulate the
divergences of quantum field theories. However, the discovery of renormalizable field theories pushed these ideas to the background until the difficulties encountered in the unification of gravity and quantum mechanics forced us to reconsider these ideas.  Indeed, strong arguments in favour of non-commutative spacetime were given much more recently in \cite{dop} and further support for non-commutative geometry came from string theory \cite{wit}. 

These observations led to a flurry of activities on non-commutative quantum field theories \cite{doug} and the possible physical consequences of non-commutative spacetime in quantum mechanics and quantum mechanical many-body systems \cite{duval}-\cite{khan}, quantum electrodynamics \cite{chai}-\cite{lia}, the standard model \cite{ohl} and cosmology  \cite{gar}, \cite{alex}. 

Despite this not much attention seems to have been paid to the formal and interpretational aspects of non-commutative quantum mechanics.  Only recently these issues were addressed in \cite{Scholtz1} where a consistent formulation and interpretation of non-commutative quantum mechanics were given. These ideas were already used in \cite{Scholtz2} to give precise meaning to the concept of a non-commutative two-dimensional well and to compute the spectrum of a particle in an infinite well.  In this article we continue to build on this analysis, but here we focus on finite wells and discuss the bound state energies and phase shifts due to scattering from a finite two-dimensional, non-commutative well. This requires a careful analysis of the matching conditions at a non-commutative boundary and the subsequent effect on the coefficients of the non-commutative wavefunction, which forms a central part of the analysis presented here. The motivation for this study is, firstly, to develop a clear understanding of the mathematical structure of the theory and, secondly, the possible physical consequences of non-commutativity, particularly in scattering data. Hopefully this can provide a guide to the analysis of more realistic three dimensional theories, which are considerably more complicated due to the breaking of rotational invariance in non-commutative theories. 

Although scattering has been studied rather extensively in the context of non-commutative quantum field theories \cite{doug, vaid}, much less has been done on potential scattering in non-commutative quantum mechanics in either two or three dimensions \cite{koch,bell,alav}.  These few studies depart from a leading order expansion in the non-commutative parameter of either the Moyal product or Bopp-shifted formulation of the Schr\"odinger equation.  This is sufficient for studying analytic potentials, but fails in the case of the well.  Furthermore, the commuting coordinates introduced in this approach do not have a clear physical meaning, which complicates the physical interpretation of these results. In contrast, the approach followed here allows an exact, albeit numerical, computation of bound state energies, phase shifts, differential and total scattering cross sections.  Furthermore the computation is interpreted within the fixed framework set out in \cite{Scholtz1}.           

To fix the notation and basic concepts, it is useful to start with a brief review of the formalism of non-commutative quantum mechanics as described in detail in \cite{Scholtz1}. The coordinates in non-commutative configuration space obey the following commutation relation,
\begin{eqnarray}
	[\hat{x},\hat{y}] = i\theta,
	\label{XYComRelation}
\end{eqnarray}
where $\theta$ is real and without loss of generality can be taken to be positive. Introducing creation and annihilation operators $b^{\dagger}$ and $b$,
\begin{eqnarray}
	\label{XhatDefinition}
	\hat{x} &=& \sqrt{\frac{\theta}{2}}\,(b + b^{\dagger}) \\
	\label{YhatDefinition}
	\hat{y} &=& \frac{\sqrt{\theta}}{i\sqrt{2}}\,(b - b^{\dagger}),
\end{eqnarray}
establishes an isomorphism between the non-commutative configuration space, ${\cal H}_c$, and boson Fock space,
\begin{eqnarray}
	{\cal H}_c = \textrm{span}\{\ket{n}\equiv \frac{1}{\sqrt{n!}}(b^{\dagger})^n\ket{0}\}^{n=\infty}_{n=0},
\end{eqnarray}
where the span is over complex numbers.  The space of physical states is represented by what we refer to as the quantum Hilbert space, ${\cal H}_q$. This space consists of the Hilbert-Schmidt operators acting on the non-commutative configuration space,
\begin{eqnarray}
	{\cal H}_q = \{\hat\psi(\hat{x},\hat{y}): \hat\psi(\hat{x},\hat{y}) \in {\cal B}({\cal H}_c),\, \textrm{tr}_c(\hat\psi(\hat{x},\hat{y})^{\dagger},\hat\psi(\hat{x},\hat{y}))<\infty\},
\end{eqnarray}
where $\textrm{tr}_c$ denotes the trace over the non-commutative configuration space and ${\cal B}({\cal H}_c)$ indicates the set of bounded operators on ${\cal H}_c$. The inner product on this space is the trace inner product 
\begin{equation}\label{inner}
	\left(\hat\phi(\hat{x}_1,\hat{x}_2),\hat\psi(\hat{x}_1,\hat{x}_2)\right) = {\rm tr_c}(\hat\phi(\hat{x}_1,\hat{x}_2)^\dagger\hat\psi(\hat{x}_1,\hat{x}_2)).
\end{equation}
We denote the elements of this space by $\hat\psi(\hat{x},\hat{y})\equiv |\hat\psi)$ and the elements of its dual (linear functionals) by $(\hat\psi|$, which maps elements of $\mathcal{H}_q$ onto complex numbers by $\left(\hat\phi|\hat\psi\right)=\left(\hat\phi,\hat\psi\right)={\rm tr_c}\left(\hat\phi(\hat{x}_1,\hat{x}_2)^\dagger\hat\psi(\hat{x}_1,\hat{x}_2)\right)$.

The non-commutative Heisenberg algebra in two dimensions reads
\begin{eqnarray}
\label{heisnc}
\left[\hat{x}_i,\hat{p}_j\right] &=& i\hbar\delta_{i,j},\nonumber\\
\left[\hat{x}_i,\hat{x}_j\right] &=& i\theta\epsilon_{i,j},\\
\left[\hat{p}_i,\hat{p}_j\right] &=& 0\nonumber.
\end{eqnarray}
A unitary representation of this algebra in terms of operators $\hat{X}_i$ and $\hat{P}_i$ acting on the quantum Hilbert space is given by
\begin{eqnarray}
\label{PxPyAction}
\hat{X}_i\hat{\psi}(\hat{x}_1,\hat{x}_2) &=& \hat{x}_i\hat{\psi}(\hat{x}_1,\hat{x}_2),\nonumber\\
\hat{P}_i\hat{\psi}(\hat{x}_1,\hat{x}_2) &=& \frac{\hbar}{\theta}\epsilon_{i,j}[\hat{x}_j,\hat{\psi}(\hat{x}_1,\hat{x}_2)],
\end{eqnarray}
i.e., the position acts by left multiplication and the momentum adjointly. We use capital letters to distinguish operators acting on quantum Hilbert space from those acting on non-commutative configuration space. It is also useful to introduce the following quantum operators (in what follows $\dagger$ refers to hermitian conjugation on the quantum Hilbert space)  
\begin{eqnarray}
	\label{qop}
	\hat{P}&=&\hat{P}_1 + i\hat{P}_2,\nonumber\\
	\hat{P}^\dagger &=& \hat{P}_1 -i\hat{P}_2.
\end{eqnarray}
We note that $\hat{P}^2=\hat{P}^2_1+\hat{P}^2_2 = P^\dagger P = PP^\dagger$.  These operators act as follow
\begin{eqnarray}
\label{PAction}
P\hat\psi(\hat{x}_1,\hat{x}_2)&=& -i\hbar \sqrt{\frac{2}{\theta}}[b,\hat\psi(\hat{x}_1,\hat{x}_2)], \\
\label{PBarAction}
P^\dagger\hat\psi(\hat{x}_1,\hat{x}_2) &=& i\hbar\sqrt{\frac{2}{\theta}}[ b^{\dagger},\hat\psi(\hat{x}_1,\hat{x}_2)].
\end{eqnarray}
This quantum system is interpreted within the usual set of axioms of commutative quantum mechanics with the simple replacement of $L^2$ by ${\cal H}_q$.  Only position measurement requires the slight, yet well established, generalization of positive operator valued measures, which we now briefly review.  

In \cite{Scholtz1} it was shown that the following operators provide a positive operator valued measure on the quantum Hilbert space
\begin{equation}
\label{povm}
\pi_z=\frac{1}{2\pi\theta}|z) e^{\stackrel{\leftarrow}{\partial_{\bar{z}}}\stackrel{\rightarrow}{\partial_z}}(z|.
\end{equation}
Here $|z)=\ket{z}\bra{z}$ with $\ket{z}$ the coherent state 
\begin{equation}
	\ket{z} = e^{-z \bar{z}/2}e^{z b^{\dagger}}\ket{0}.
\end{equation}
The probability of finding the particle at position $\left(x_1,x_2\right)$, given that the system is described by the pure state density matrix $\rho=|\hat\psi )(\hat\psi|$ is then
\begin{eqnarray}
	P(x_1,x_2)={\rm tr_q}\left(\pi_z\rho\right)=\left(\hat\psi|\pi_z|\hat\psi\right)=\frac{1}{2\pi\theta}\left(\hat\psi|z\right) e^{\stackrel{\leftarrow}{\partial_{\bar{z}}}\stackrel{\rightarrow}{\partial_z}}\left(z|\hat\psi\right).
\end{eqnarray}
This naturally leads us to interpret $\left(z|\hat\psi\right)=\langle z|\hat\psi|z\rangle$ as the wavefunction of the non-commutative system.

The paper is organised as follows.  We start by giving a brief description of the construction of a non-commutative well. We then move on to the non-commutative wavefunction and explain how one connects the different parts of the wavefunction at the boundary of the well.  An important section then follows on the relationship between the Fock space representation and the wavefunction. We then turn to some applications of the non-commutative formulation of quantum mechanics by calculating some bound states and phase shifts as well as the total cross-section for the non-commutative well.

\section{The Non-commutative Well}
One of the most famous statements in physics is undoubtably Heisenberg's Uncertainty Principle. It prevents you from knowing the simultaneous position and momentum of a particle with arbitrary accuracy. This stems from the fact that, in quantum physics, position and momentum no longer commute. This leads to an interesting complication when doing non-commutative quantum mechanics: how does one define a non-commutative well?

Since the $x$- and $y$-coordinates no longer commute, we see that, in the spirit of the Uncertainty Principle, we can no longer specify that the potential, as a function of the coordinates, has a certain value at a specific boundary.

We can solve the problem of defining a potential well in a non-commutative space by using projection operators \cite{Scholtz2}. Using the mathematical tools described in \cite{Scholtz2}, each projection operator specifies one region of constant potential. Therefore, in the case of a single well, we would have,
\begin{eqnarray}
	\label{PDef}
	P &=& \sum^N_{n=0}\ket{n}\bra{n} \\
	\label{QDef}
	Q &=& \sum^{\infty}_{n=N+1}\ket{n}\bra{n} \\
	Q &=& 1-P.
	\label{ProjectionOp}
\end{eqnarray}
Using these projection operators a piecewise constant potential (here a well) is simply defined as,
\begin{equation}
	\hat{V} = V_1\,P + V_2\,Q,
\end{equation}
where $V_1$ and $V_2$ are constants.

We see that if we define $\hat{r}^2 \equiv \hat{x}^2 + \hat{y}^2 = \theta(2b^{\dagger}b + 1)$ the radius of the well is defined by $N$ according to $R^2 = \theta(2N + 1)$. It is immediately clear from this expression that the radius (and hence also area) of the potential is quantised in units of $\theta$.  This quantisation of the area of the potential in terms of $\theta$ is, of course, what drives any departure from normal commutative physics. However, it is important to note that this effect becomes weaker as $\theta$ becomes smaller and should go over to the commutative case in the limit of $\theta \rightarrow 0$. For the potential well this has already been checked in \cite{Scholtz2} and this article will focus more on some of the departures from the commutative senario for finite values of $\theta$.

As with its commutative counterpart, the non-commutative potential well is a conceptually simple system, which already allows us to investigate a number of differences that non-commutativity brings. The examples discussed in this article include how bound states are influenced and changes to the phase shifts during time-independent scattering.

\subsection{Wavefunction}
\label{WavefunctionSection}
The wavefunction in a constant potential obeys the following Schr\"{o}dinger equation,
\begin{equation}
	\hat{P}^2\hat{\psi} = k^2 \hbar^2 \hat{\psi},
	\label{NCConstantSchrodinger}
\end{equation}
where $k^2 = 2\mu (E-V)/\hbar^2$, $V$ constant. 

As elaborated in \cite{Scholtz1} and the discussion above, the quantity $\rbraket{z}{\hat\psi}=\langle z|\hat\psi|z\rangle$ plays the role of the wavefunction.  As such it plays a central role in the scattering theory developed later and we would like to find the differential equation obeyed by it.  We therefore construct
\begin{equation}
	\bra{z}\hat{P}^2\hat{\psi}\ket{z} = k^2\hbar^2\bra{z}\hat{\psi}\ket{z} = k^2\hbar^2\rbraket{z}{\hat{\psi}}.
	\label{NCSchrodingerPosistionRep}
\end{equation}
Consider the LHS of \eqref{NCSchrodingerPosistionRep}. From the definition of the momentum operator \cite{Scholtz1,Scholtz2},
\begin{eqnarray}
	\bra{z}\hat{P}^2\hat{\psi}\ket{z} &=& \frac{2\hbar^2}{\theta}\bra{z}[b^{\dagger},[b,\hat{\psi}]]\ket{z}.
\end{eqnarray}
To write this as a differential equation we use the following identities 
\begin{eqnarray}
	b\ket{z} &=& z\ket{z},\\
	\langle z|b^\dagger &=& \bar{z}\langle z|,\\
 b^{\dagger}\ket{z} &=& \left(\pafg{}{z} + \frac{1}{2}\bar{z}\right)\ket{z},\\
 	\bra{z}b &=& \bra{z}\left(\stackrel{\leftarrow}{\pafg{}{\bar{z}}} + \frac{1}{2}z\right).
\end{eqnarray}
where the arrow over the partial derivative indicates differentiation to the left. After some algebra one obtains
\begin{eqnarray}
	\bra{z}b^{\dagger}[b,\hat{\psi}] - [b,\hat{\psi}] b^{\dagger}\ket{z} &=& -\pafg{}{z}\pafg{}{\bar{z}}\bra{z}\hat{\psi}\ket{z} = -\pafg{}{z}\pafg{}{\bar{z}}\rbraket{z}{\hat\psi}.
\end{eqnarray}
We therefore find that the non-commutative wavefunction obeys
\begin{eqnarray}
	-\pafg{}{z}\pafg{}{\bar{z}}\rbraket{z}{\hat\psi} = \frac{\theta k^2}{2}\rbraket{z}{\hat\psi}.
\end{eqnarray}
Putting $\rbraket{z}{\hat\psi} = \psi(z,\bar{z})$ and transforming to variables $x$ and $y$, where $z = x + iy$, leads to
\begin{eqnarray}
	-(\pafg{^2}{x^2} + \pafg{^2}{y^2})\psi(x,y) = 2\theta k^2\psi(x,y),
\end{eqnarray}
which is the standard Schr\"{o}dinger equation.

At this point it would be natural to wonder why one would even bother with non-commutativity if the wavefunction is the same as in the commutative case. Indeed, it is well known (see e.g. \cite{gov}) that introducing non-commutativity changes nothing for a completely free particle. However, in the presence  of a potential non-commutativity has physical consequences and in particular for the non-commutative well these differences arise from the matching conditions that need to be satisfied at the boundary of the two (or more) regions of constant potential, which are quite different from the commutative case.  We discuss these matching conditions in the next section.

\subsubsection{Matching Conditions}
\label{MatchingConditionsSection}
Finding the matching conditions of the wavefunctions in the different regions of the non-commutative well starts out similarly to the commutative case. Here the Schr\"{o}dinger equation is given by
\begin{equation}
	\frac{\hat{P}^2}{2\mu}\hat{\psi} + (P V_1 + Q V_2)\hat{\psi} = E\hat{\psi}.
	\label{NCSchrodingerFull}
\end{equation}
We construct two solutions of the Schr\"{o}dinger equation at the same energy for the two different values of the potential,
\begin{eqnarray}
	\label{NCSchrodingerV1}
	\frac{\hat{P}^2}{2\mu}\hat{\psi}_1 + V_1\hat{\psi}_1 = E\hat{\psi}_1 \\
	\label{NCSchrodingerV2}
	\frac{\hat{P}^2}{2\mu}\hat{\psi}_2 + V_2\hat{\psi}_2 = E\hat{\psi}_2.
\end{eqnarray}
However, instead of requiring that the solutions and their derivatives match at the boundary, we require that a consistency equation is satisfied. The details of deriving this equation can be found in \cite{Scholtz2}.
If we suppose that the full solution $\hat{\psi}$ is given by
\begin{equation}
	\hat{\psi} = P\hat{\psi}_1 + Q\hat{\psi}_2,
	\label{NCSchrodingerFullSol}
\end{equation}
and we substitute \eqref{NCSchrodingerFullSol} into \eqref{NCSchrodingerFull} we obtain
\begin{eqnarray}
	\frac{\hat{P}^2}{2\mu}(P\hat{\psi}_1 + Q\hat{\psi}_2) + (V_1 P\hat{\psi}_1 + V_2 Q\hat{\psi}_2) = E P\hat{\psi}_1 + E Q\hat{\psi}_2.
\end{eqnarray}
Setting $\Omega = \left[\hat P^2,P\right]$ and multiplying (\ref{NCSchrodingerV1}) by $P$ and (\ref{NCSchrodingerV2}) by $Q$ from the left shows that (\ref{NCSchrodingerFullSol}) is a solution provided that 
\begin{equation}
	\Omega\hat{\psi}_1 = \Omega\hat{\psi}_2.
	\label{ConsistencyEq}
\end{equation}

This is the consistency equation that determines the matching conditions. By using the definition \eqref{PDef} of $P$, the action of $b$ and $b^{\dagger}$ on the harmonic oscillator states and taking the inner product of $\Omega\hat{\psi}_1$ and $\Omega\hat{\psi}_2$ with $\bra{N}$ and $\ket{\ell}$, it has been shown in \cite{Scholtz2} that \eqref{ConsistencyEq} is written more suggestively as
\begin{eqnarray}
	\label{GeneralMatchingConditions1}
	\bra{N+1}\hat{\psi}_1\ket{\ell+1} &=& \bra{N+1}\hat{\psi}_2\ket{\ell+1} \\
	\label{GeneralMatchingConditions2}
	\bra{N}\hat{\psi}_1\ket{\ell-1} &=& \bra{N}\hat{\psi}_2\ket{\ell-1}.
\end{eqnarray}

For subsequent discussions where we calculate things numerically we typically look at sectors with fixed angular momentum. In \cite{Scholtz2} the form of $\hat{\psi}$ for a specific value of angular momentum, namely $\hat{\psi}_m$, was touched on very briefly. We now explicitly construct the eigenstates $\hat{\psi}_m$ of the angular momentum operator. By considering \cite{Scholtz1} the operator that generates rotations around the $z$-axis, the angular momentum operator was derived as
\begin{equation}
	\hat{L}_z = \sqrt{\frac{\theta}{2}}(b+b^{\dagger})\hat{P}_y + i\sqrt{\frac{\theta}{2}}(b-b^{\dagger})\hat{P}_x + \frac{\theta}{2\hbar}\hat{P}^2.
	\label{AngularMomentumOperator}
\end{equation}
We can write the operator $\hat{\psi}$ generally as (see \cite{Scholtz2})
\begin{eqnarray}
	\hat{\psi} = \sum^{\infty}_{k=0}\sum^{\infty}_{\ell=0}a_{k,\ell}(b^{\dagger})^k b^{\ell} = \sum^{\infty}_{m=-\infty}\hat{\psi}_m,
\end{eqnarray}
where
\begin{eqnarray}
	\hat{\psi}_m = \sum^{\infty}_{k=0}a_{k,k+m}(b^{\dagger})^k b^{k+m},\;\;m\geq0.
\end{eqnarray}
Using \eqref{PxPyAction}, \eqref{PAction} and \eqref{PBarAction} we can calculate the action of the angular momentum operator on $\hat{\psi}_m$:
\begin{eqnarray}
	\hat{L}_z\,\hat{\psi}_m = -\hbar[b^{\dagger}b,\hat{\psi}_m] = \hbar m\,\hat{\psi}_m.
\end{eqnarray}
We therefore conclude that the $\hat{\psi}_m$ correspond to partial waves with angular momentum $m$.  For these $\hat{\psi}_m$ the matching conditions become,
\begin{eqnarray}
	\label{AngularMatchingConditions1}
	\bra{N+1}\hat{\psi}_{1,m}\ket{N+m+1} &=& \bra{N+1}\hat{\psi}_{2,m}\ket{N+m+1} \\
	\label{AngularMatchingConditions2}
	\bra{N}\hat{\psi}_{1,m}\ket{N+m} &=& \bra{N}\hat{\psi}_{2,m}\ket{N+m},
\end{eqnarray}
and these will be the ones used predominantly during numerical calculations.

On a side note, in the discussion leading up to here we worked under the assumption that $m$ is positive or zero. The mathematics in the case where $m < 0$ is the same as that of positive $m$ and is not discussed in this article. One difference worth mentioning though is that, unlike positive $m$, there is a cutoff for negative $m$ in the form of $|m| \leq N$ \cite{Scholtz2}. This asymmetry between $m > 0$ and $m < 0$ is caused by an implied parity violation in the commutator of $\hat{x}$ and $\hat{y}$ due to the choice of the sign of $\theta$ and the fact that the commutator is not symmetric under parity transformations. The asymmetry can also be understood from the point of view of time reversal symmetry breaking as discussed in \cite{Scholtz1}.

\subsubsection{Relation between $\bra{z}\hat{\psi}\ket{z}$ and $\bra{n}\hat{\psi}\ket{n+m}$}
\label{ConnectionBetweenZandNSection}
In previous sections we have seen two important things, namely that the non-commutative wavefunction, $\rbraket{z}{\hat{\psi}}$, obeys a normal commutative type Schr\"{o}\-dinger equation and that the correct manner in which to match the various parts of $\hat{\psi}$ at potential boundaries is to use the matching conditions in terms of the oscillator basis states. Knowing that the non-commutative wavefuntion is essentially a plane wave when the potential is constant and that we can use all the standard tools such as partial wave expansions is of course very useful when we want to calculate phase shifts, for example. However, as we have just seen, the coefficients of these partial waves will all come from the matching conditions which are given in the oscillator basis. It is therefore crucial that we can convert what we know in the oscillator basis into the position basis.

To find how the two bases on classical configuration space are related to each other we start by looking at the expansion of a general operator over the coherent states \cite{Klauder},
\begin{equation}
	\hat{A} = \int {\rm d}z\, a(z) \ket{z}\bra{z}.
\end{equation}
With some algebra it can be shown that $a(z)$ can be written as
\begin{equation}
	a(z) = \frac{1}{\pi}e^{-\frac{\partial^2}{\partial z \partial\bar{z}}}\bra{z}\hat{A}\ket{z}.
\end{equation}
Taking $\hat{A} = \hat{\psi}_m$ we see that the above equation simplifies dramatically, since $\bra{z}\hat{\psi}_m\ket{z}$ is an eigenfunction of the $-\frac{\partial^2}{\partial z \partial\bar{z}}$ operator as we saw in Section \ref{WavefunctionSection}. Therefore,
\begin{eqnarray}
	-\frac{\partial^2}{\partial z \partial\bar{z}}\bra{z}\hat{\psi}_m\ket{z} = \frac{\theta k^2}{2} \bra{z}\hat{\psi}_m\ket{z} \\
	\Rightarrow \hat{\psi}_m = \frac{1}{\pi} e^{\frac{\theta k^2}{2}}\int {\rm d}z\, \bra{z}\hat{\psi}_m\ket{z}\ket{z}\bra{z}.
\end{eqnarray}
Taking then the general matrix element in the oscillator basis gives
\begin{eqnarray}
	\bra{n}\hat{\psi}_m\ket{\ell} &=& \frac{1}{\pi} e^{\frac{\theta k^2}{2}}\int{\rm d}z\,\bra{z}\hat{\psi}_m\ket{z}\braket{n}{z}\braket{z}{\ell}.
\end{eqnarray}
Transforming $\bra{z}\hat{\psi}_m\ket{z}$ into polar co-ordinates, $z = re^{i\phi}$, and using $\braket{n}{z} = e^{-\frac{|z|^2}{2}}\frac{z^n}{\sqrt{n!}}$ gives us
\begin{eqnarray}
	\bra{n}\hat{\psi}_m\ket{\ell} &=& \frac{1}{\pi} e^{\frac{\theta k^2}{2}}\int{\rm d}\phi\,{\rm d}r r\,e^{im\phi}(A\,J_m(\sqrt{2\theta}k r)+ B\,Y_m(\sqrt{2\theta}k r)) \nonumber\\
	& & \times e^{-r^2}\frac{r^n e^{in\phi}}{\sqrt{n!}}\frac{r^{\ell}e^{-i\ell\phi}}{\sqrt{\ell!}} \\
	&=& \frac{e^{\frac{\theta k^2}{2}}}{\pi\sqrt{n!}\sqrt{\ell!}}\underbrace{\int^{2\pi}_{0}{\rm d}\phi\,e^{i(m +n-\ell)\phi}}_{2\pi\delta_{\ell,n+m}}\int^{\infty}_{0}{\rm d}r\, r^{n+\ell+1}e^{-r^2} \nonumber\\
	& & \times (A\,J_m(\sqrt{2\theta}k r)+ B\,Y_m(\sqrt{2\theta}k r)) \\
	\Rightarrow\bra{n}\hat{\psi}_m\ket{n+m} &=& \frac{2e^{\frac{\theta k^2}{2}}}{\sqrt{n!}\sqrt{(n+m)!}}\int^{\infty}_{0}{\rm d}r\, r^{2n+m+1}e^{-r^2} \nonumber \\
	& & \times(A\,J_m(\sqrt{2\theta}k r)+ B\,Y_m(\sqrt{2\theta}k r))
\end{eqnarray}
The integral over $J_m$ is given in \cite{Gradshteyn},
\begin{equation}
	\int^{\infty}_{0}{\rm d}r\, r^{2n+m+1}e^{-r^2} J_m(\sqrt{2\theta}k r) = \frac{n!}{2}e^{-w}w^{m/2}L^m_n(w),
\end{equation}
where $w = \frac{\theta k^2}{2}$.\\
Doing the integral over $Y_m$ is slightly more involved. We first look at the integral for $Y_{\nu} = \frac{1}{\sin(\nu\pi)}(\cos(\nu\pi)J_{\nu} - J_{-\nu})$ where $\nu \in \cal{R}$:
\begin{eqnarray}
	& & \int^{\infty}_{0}{\rm d}r\, r^{2n+\nu+1}e^{-r^2} Y_{\nu}(\sqrt{2\theta}k r) \nonumber\\
	&=& \frac{1}{\sin(\nu\pi)}\left[\cos(\nu\pi)\underbrace{\frac{w^{\nu/2}}{2}\frac{\Gamma(n+\nu+1)}{\Gamma(\nu+1)}M(n+\nu+1,\nu+1,-w)}_{(a)}\right. \nonumber\\
	& & - \left.\underbrace{\frac{w^{-\nu/2}}{2}\frac{\Gamma(n+1)}{\Gamma(1-\nu)}M(n+1,1-\nu,-w)}_{(b)}\right],
\end{eqnarray}
where $M(a,b,x)$ is the first solution of the confluent hypergeometric differential equation.
The substitutions (a) and (b) we have made required the evaluation of the following integrals, the solutions of which can be found in \cite{Gradshteyn}:
\begin{eqnarray}
	\textrm{(a):}& & \int^{\infty}_{0}{\rm d}r\, r^{2n+\nu+1}e^{-r^2} J_{\nu}(\sqrt{2\theta}k r),\\
	\textrm{(b):}& & \int^{\infty}_{0}{\rm d}r\, r^{2n+\nu+1}e^{-r^2} J_{-\nu}(\sqrt{2\theta}k r),
\end{eqnarray}
We can now write our integral over $Y_{\nu}$ as
\begin{eqnarray}
	& & \int^{\infty}_{0}{\rm d}r\, r^{2n+\nu+1}e^{-r^2} Y_{\nu}(\sqrt{2\theta}k r) \nonumber\\
	&=& -\frac{w^{-\nu/2}}{2\pi}\Gamma(n+\nu+1)\Gamma(n+1)\frac{\pi}{\sin(\nu\pi)}e^{-w}\left[\frac{M(-n-\nu,1-\nu,w)}{\Gamma(n+\nu+1)\Gamma(1-\nu)}\right.\nonumber \\
	& & \left.- \frac{e^{i\nu\pi}+e^{-i\nu\pi}}{2}w^{\nu}\frac{M(-n,\nu+1,w)}{\Gamma(n+1)\Gamma(\nu+1)}\right] \\
	&=& -\frac{w^{-\nu/2}}{4\pi}\Gamma(n+\nu+1)\Gamma(n+1)\frac{\pi}{\sin(\nu\pi)}e^{-w}\nonumber\\
	& & \times\left[\frac{M(-n-\nu,1-\nu,w)}{\Gamma(n+\nu+1)\Gamma(1-\nu)} - e^{i\nu\pi}w^{\nu}\frac{M(-n,\nu+1,w)}{\Gamma(n+1)\Gamma(\nu+1)} \right.\nonumber \\
	& & +\left.\frac{M(-n-\nu,1-\nu,w)}{\Gamma(n+\nu+1)\Gamma(1-\nu)} - e^{-i\nu\pi}w^{\nu}\frac{M(-n,\nu+1,w)}{\Gamma(n+1)\Gamma(\nu+1)}\right]\\
	&=& -\frac{w^{-\nu/2}}{2\pi}\Gamma(n+\nu+1)\Gamma(n+1)\nonumber \\
	& & \times\left[\frac{1}{2}U(n+1,1-\nu,w e^{i\pi})+ \frac{1}{2}U(n+1,1-\nu,w e^{-i\pi})\right],
\end{eqnarray}
where we used $M(a,b,x) = e^x M(b-a,b,-x)$ in the first line and the analytic continuation of $U$ to get to the last line (see \cite{Abramowitz}).  The notation $we^{\pm i\pi}$ denotes evaluation just above ($+$) or just below $(-)$ the negative real axis, which is the branch cut for the function $U$. For real values of $w$ the sum of the $U$'s in the last line simplifies to ${\rm Re}[U(n+1,1-\nu,-w)]$. $U$ is defined even in the limit of $\nu \rightarrow m$, where $m$ is integer, so this also solves our original integral for $Y_m$. Putting everything together we find
\begin{eqnarray}
	\bra{n}\hat{\psi}_m\ket{n+m} &=& A\frac{\sqrt{n!}}{\sqrt{(n+m)!}}w^{m/2}L^m_n(w)\nonumber\\
	 && \!\!\!\!\!\!\!\!\!\! -\frac{B}{\pi}\sqrt{n!(n+m)!}\,e^{w}w^{-m/2}{\rm Re}[U(n+1,1-m,-w)].
	 \label{WaveCoefficients}
\end{eqnarray}

We remark that in \cite{Scholtz2} the matrix elements $\bra{n}\hat{\psi}_m\ket{n+m}$ were solved using recursion relations. The result obtained was
\begin{eqnarray}
	\bra{n}\hat{\psi}_m\ket{n+m} &=& c_1(m,-w)\sqrt{\frac{m!n!}{(n+m)!}}L^m_n(w)\nonumber \\
	& & \!\!\!\!\!\!\! + c_2(m,-w)\sqrt{\frac{n!(n+m)!}{m!}}U(n+1,1-m,-w).
	\label{MatrixCoefficients}
\end{eqnarray}
This result is actually only valid when $U(n+1,1-m,-w)$ is real, i.e. $w < 0$, as is the case for bound states studied in reference \cite{Scholtz2}. For scattering states the correct form to use is \eqref{WaveCoefficients}. In the former case we therefore have a simple way of relating the two sets of coefficients, $c_1$ and $c_2$ and $A$ and $B$ by comparing \eqref{WaveCoefficients} and \eqref{MatrixCoefficients}:
\begin{eqnarray}
	\label{AconnectionC1}
	A &=& \sqrt{m!}\,w^{-m/2}c_1(m,-w) \\
	\label{BconnectionC2}
	B &=& -\frac{\pi}{\sqrt{m!}} w^{m/2}e^{-w} c_2(m,-w)
\end{eqnarray}

\subsection{Bound State Energies}
In this section we come to the first application of the non-commutative ideas described in this article. To find the bound states in the commutative case one would first solve the Schr\"{o}dinger equation in the inner and outer regions of the well. Inside one obtains a linear combination of the Bessel functions $J_m$ and $Y_m$. Realizing that $Y_m$ is singular at the origin one then sets its coefficient to zero. Similarly, outside one has a linear combination of $I_m$ and $K_m$, but due to the exponentially growing nature of $I_m$ its coefficient is also chosen as zero. In the non-commutative case we have seen that the general form of $\bra{n}\hat{\psi}_m\ket{n+m}$ is a linear combination of a Laguerre polynomial $L^m_n$ and a confluent hypergeometric function $U$. Since the non-commutative wavefunction has the same form as it does commutatively, namely a linear combination of $J_m$ and $Y_m$ inside the well and of $I_m$ and $K_m$ outside the well, we see that the same restrictions to the coefficients apply in the non-commutative case. By looking at the derivation of \eqref{WaveCoefficients} we see that these restrictions imply that the coefficient of $U$ inside the well and the coefficient of $L^m_n$ outside the well must be zero. The same result can be found by looking at the commutative limit of $\bra{n}\hat{\psi}_m\ket{n+m}$ (see \cite{Scholtz2}). 

In the commutative case one would then match up $J_m$ and $K_m$ and their derivatives at the boundary and solve for the energy. Using what we have just said about the coefficients in the non-commutative case, the matching conditions \eqref{AngularMatchingConditions1} and \eqref{AngularMatchingConditions2} reduce to
\begin{eqnarray}
	\frac{L^m_{N+1}(\theta E)}{L^m_{N}(\theta E)} = (N+m+1)\frac{U(N+2,1-m,\theta(V-E))}{U(N+1,1-m,\theta(V-E))},
	\label{NCBoundStateMatchingCondition}
\end{eqnarray}
where we have taken $\mu = \hbar = 1$ and divided \eqref{AngularMatchingConditions1} by \eqref{AngularMatchingConditions2}. This equation then gives us the bound state engergies for a non-commutative well of depth $V$ in the angular momentum sector $m$. In this article, these energies were found by graphically inspecting the LHS and RHS of \eqref{NCBoundStateMatchingCondition} and then searching numerically near these values.

\begin{figure}[tb]
\begin{center}
\epsfig{figure=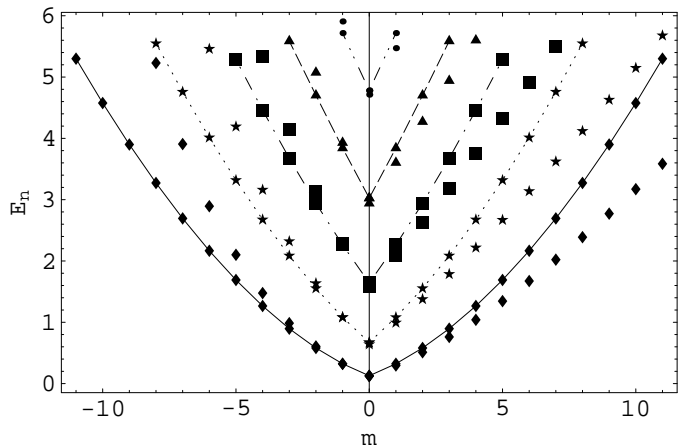,height=6cm}
\end{center}
\caption[NC and commutative bound state energies]{This figure shows the commutative and non-commutative bound state energies for a well of depth $V=6$ with a radius of $\sqrt{20}$. $N$ was chosen as 10. Symbols which are the same indicate the same energy level, i.e. diamonds indicate the ground state, stars the first excited state, etc.. Connected symbols are the commutative energies and unconnected ones indicate the non-commutative energies.}
\label{BoundStatesFigure1}
\end{figure}

In Figure \ref{BoundStatesFigure1} we show some numerical results for both positive and negative angular momentum. We see that the energy of a positive $m$ bound state is always lower than the corresponding commutative one and that the opposite is true for negative $m$. This leads to there being more bound states for some $m$ than is the case commutatively. Once again the opposite is true for $m < 0$. One thing to keep in mind is that $\theta$ is large ($\theta = \frac{20}{21}$) in this calculation. For realistic values of $\theta$, which are extremely small, the split in energy between positive and negative $m$ becomes very small as well and hence also the chance of seeing a lost or gained bound state. The graph also shows that non-commutative bound state energies differ more and more from their commutative counterparts as $|m|$ increases in relation to $N$. Choosing $N$ too large will therefore not always allow us to see where the non-commutative changes come in. As we mentioned very briefly in Section \ref{MatchingConditionsSection} $|m|$ cannot be larger than $N$ for negative $m$ and this is clearly visible in Figure \ref{BoundStatesFigure1}.

\subsection{Scattering}
In this section on scattering, we considered only time-independant scattering. In the part that covers phase shifts we briefly describe how the phase shifts were calculated and then show some results. Once we have done that we use the calculated phase shifts to find the total and differential scattering cross-section.

\subsubsection{Phase Shifts}
Let us look briefly at the commutative case. Since the range of most scattering potentials is finite (which is certainly the case for a well), a particle being scattered from such a potential is essentially free most of the time. If we were to split the particle's wavefunction into an incoming and outgoing part, we could include the effect of a scattering potential on the particle by adding a phase shift to the outgoing wave. The reason it is only a phase shift is to preserve unitarity. In the case of a well, however, we can of course solve the Schr\"{o}dinger equation everywhere. We can then compare this exact solution to the free particle plus phase shift wavefunction at large distances from the potential and obtain an equation for the phase shift in terms of the coefficients of the exact solution. In terms of radial co-ordinates, the exact wavefunction outside the well (for a specific $m$) is given by,
\begin{equation}
	\psi_m = A J_m(k r) + B Y_m(k r),
\end{equation}
where $k^2 = \frac{2\mu}{\hbar^2}(E-V)$. All this then leads to the well known equation for the phase shift $\delta_m$
\begin{equation}
	\tan\delta_m = -\frac{B}{A}.
	\label{PhaseShift}
\end{equation}

One important message this article tries to convey is that we are constantly working with two parallel pictures when dealing with non-commuta\-tivity. On the one hand we have the have the position representation $\left(z|\hat\psi\right)$ of $\hat{\psi}$, which is very similar mathematically to the commutative case and helps us think about how certain calculations should be done in the non-commutative case, since the analogy is so clear. On the other hand we have the description of $\hat{\psi}$ in the oscillator basis which differs substantially from how one does things commutatively, but is the natural basis in the description of the well. So, whereas one would find $A$ and $B$ by matching the wavefunctions and their derivatives inside and outside the well at the boundary in the commutative case, we have a two-step process in the non-commutative case. The first step involves the position representation, where we realise that the non-commutative wavefunction has the same form as is the case commutatively and we can also make a partial wave expansion. This allows us to follow the same arguments and to arrive at the same equation \eqref{PhaseShift} for the phase shifts. The second step involves the oscillator basis, where we can explicitly calculate the coefficients $A$ and $B$ from the non-commutative matching conditions (\eqref{AngularMatchingConditions1}, \eqref{AngularMatchingConditions2}) and equation \eqref{WaveCoefficients} derived in Section \ref{ConnectionBetweenZandNSection}.

\begin{figure}[tb]
\begin{center}
\subfigure[\label{PhaseShiftFigureSub1}]{\epsfig{figure=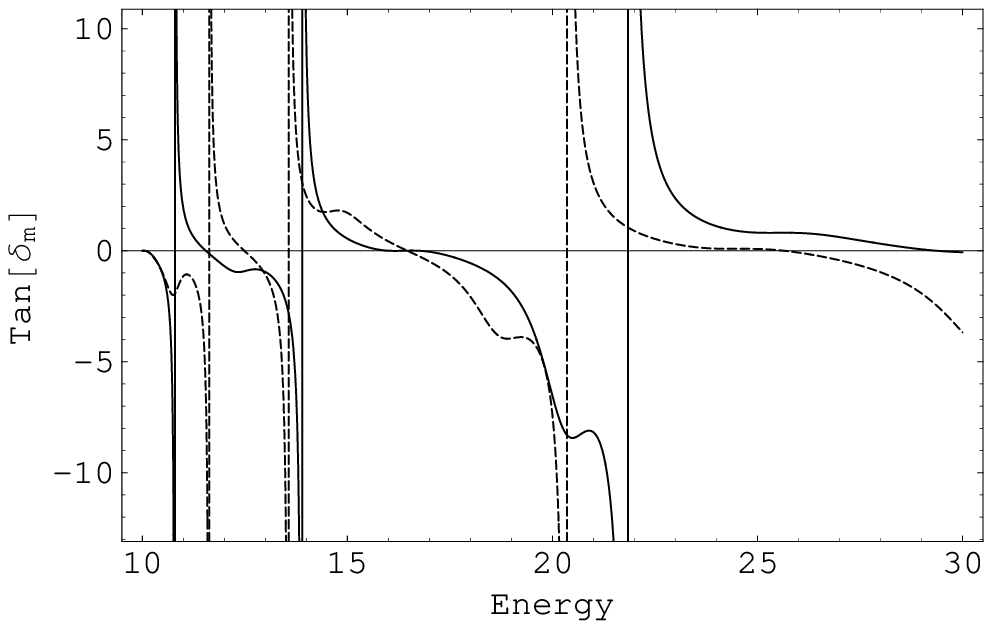,height=4cm}}
\subfigure[\label{PhaseShiftFigureSub2}]{\epsfig{figure=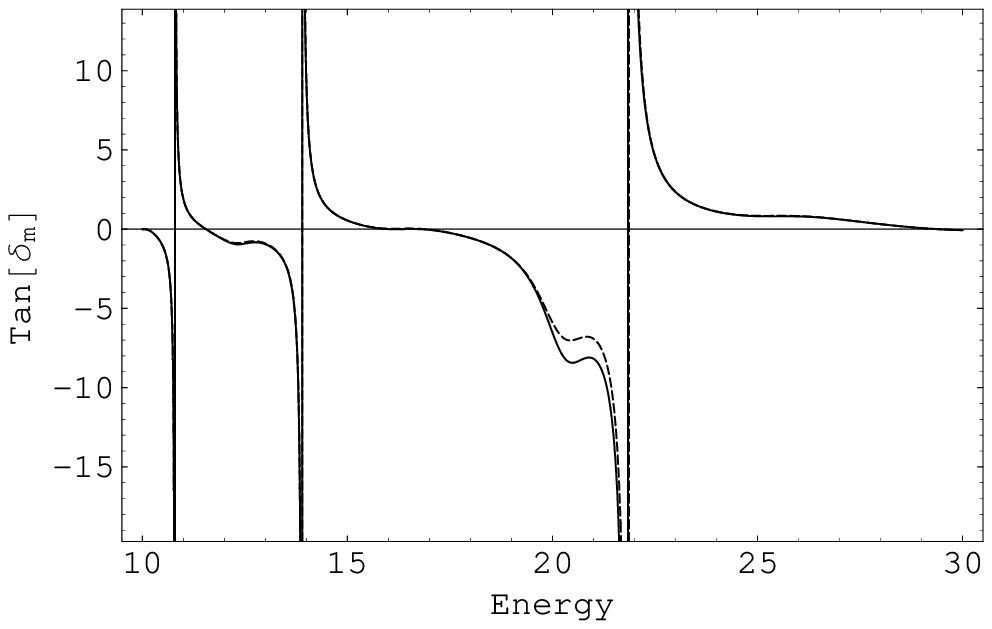,height=4cm}}
\subfigure[\label{PhaseShiftFigureSub3}]{\epsfig{figure=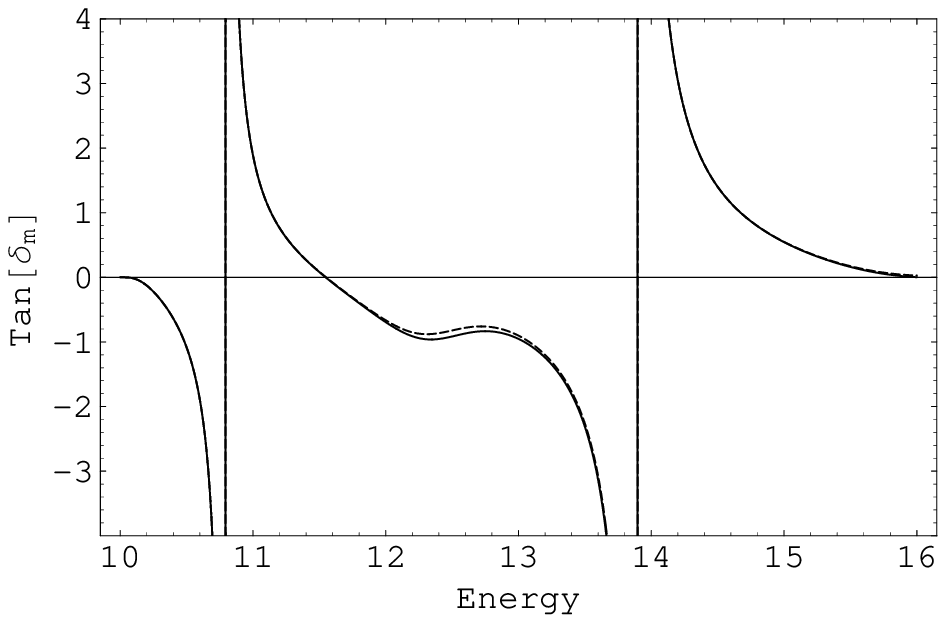,height=4cm}}
\end{center}
\caption[Tangent of NC and commutative phase shifts]{Tangent of the phase shift for the commutative (solid line) and non-commutative (dashed line) cases in the $m=4$ sector. In all three figures, well height was $V=10$ and the radius was $R=\sqrt{20}$. Figure (a) and (b) are the same except that in the former case $N=10$ ($\theta = \frac{20}{21}$) and in the latter $N=1000$ ($\theta = \frac{20}{2001}$). Figure (c) is a close up view of of Figure (b) over the first six units of energy.}
\end{figure}

The results of such a calculation are shown in Figures \ref{PhaseShiftFigureSub1} and \ref{PhaseShiftFigureSub2}. As expected the correspondence between commutativity and non-commutativity improves as $\theta$ becomes smaller.  There are however some deviations near $E = 13$ in Figure \ref{PhaseShiftFigureSub3} and near $E = 21$ in Figure \ref{PhaseShiftFigureSub2} (though these are not the only places where this occurs). These figures would then seem to suggest that there are points where convergence of the non-commutative case to the commutative one is slower. The physical origin of these large deviations in the non-commutative case is not yet understood, but it is hoped that further investigation will reveal the underlying reason as well as pointing to a concrete place to search for non-commutativity in an experiment.

\subsubsection{Total Scattering Cross-section}
The formula for the differential scattering cross-section in two dimensions is very similar in both form and derivation to that of three dimensions and we will not go into detail (see \cite{Adhikari}) here. For our purposes, we will simply use the following (commutative) formulas,
\begin{eqnarray}
	\afg{\sigma}{\phi} &=& \frac{1}{k}|f_k(\phi)|^2 \\
	f_k(\phi) &=& \sqrt{\frac{2}{\pi}}\sum^{\infty}_{m=0}\epsilon_m\cos(m\phi)e^{i\delta_m}\sin(\delta_m)\;\;\;\textrm{where}\;\;\;\epsilon_0 = 1,\epsilon_{m\ge1}=2 \\
	\sigma &=& \int^{2\pi}_{0}\sigma(\phi){\rm d}\phi = \frac{4}{k}\sum^{\infty}_{m=0}\epsilon_m\sin^2(\delta_m),
\end{eqnarray}
where $k^2 = \frac{2\mu}{\hbar^2}(E-V)$, $\afg{\sigma}{\phi}$ is the differential scattering cross-section, $f_k(\phi)$ is the scattering amplitude and $\sigma$ is the total scattering cross-section.
\begin{figure}[tb]
\begin{center}
\subfigure[\label{TDSCFigure1}]{\epsfig{figure=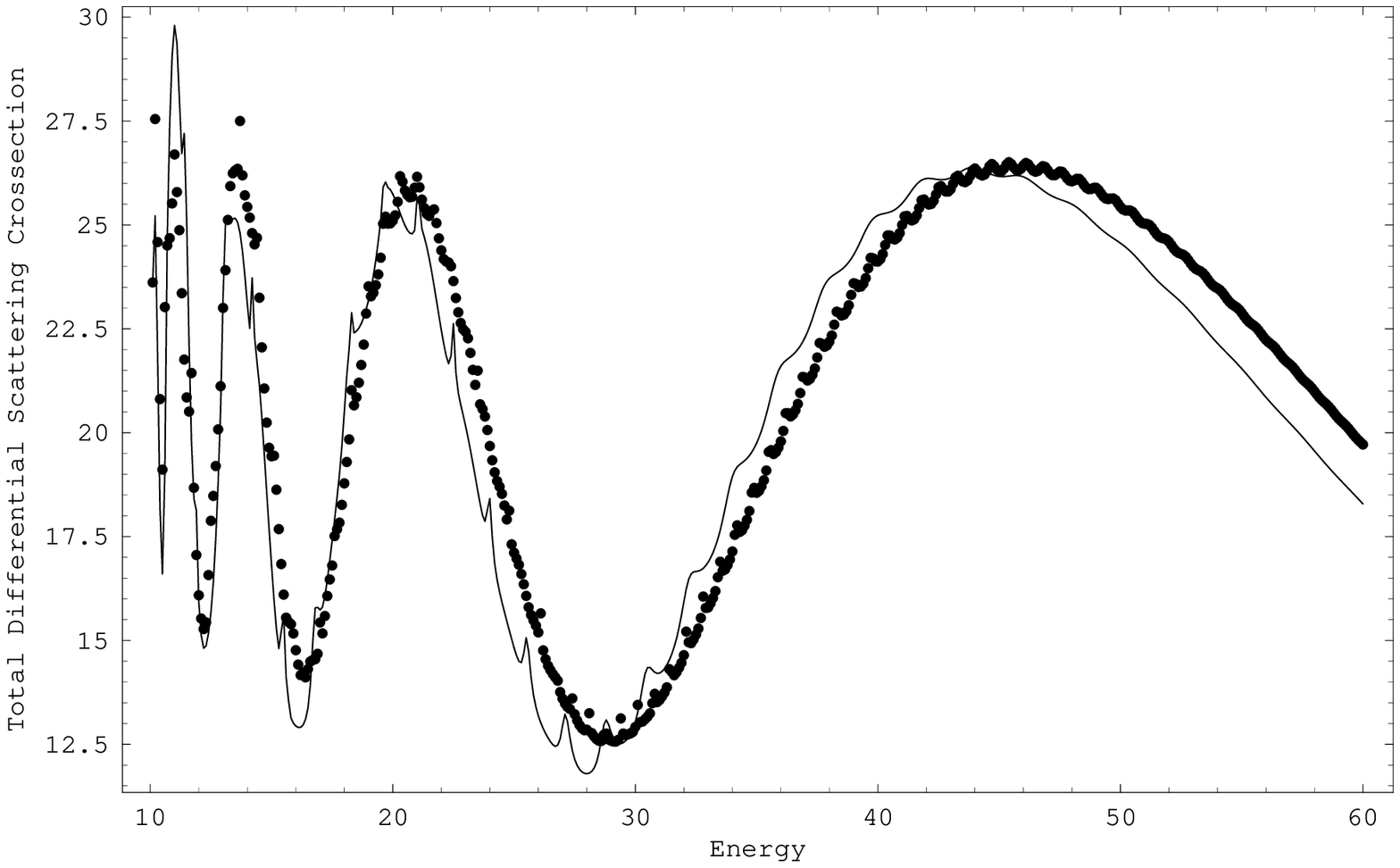,height=4.2cm}}
\subfigure[\label{TDSCFigure2}]{\epsfig{figure=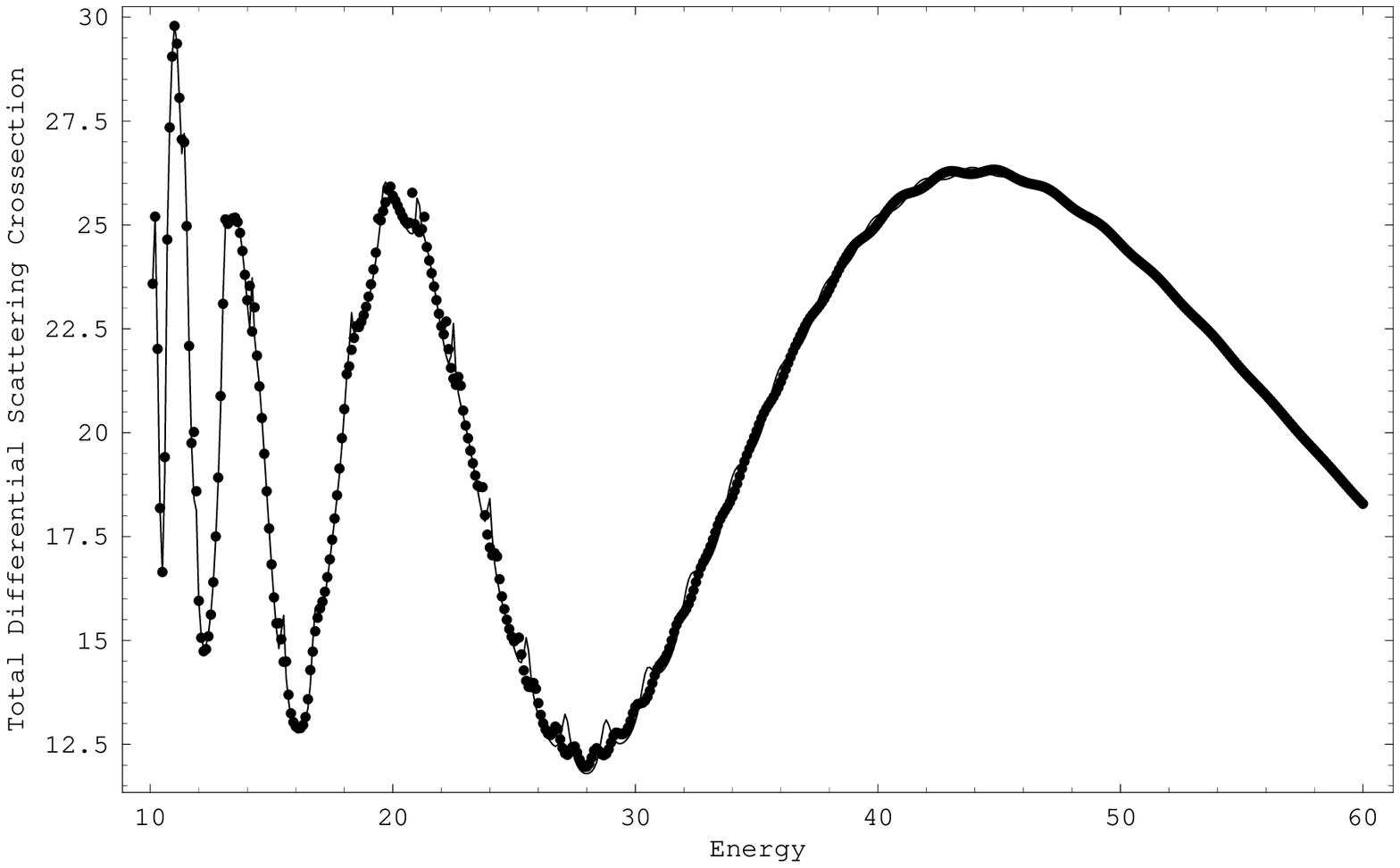,height=4.2cm}}
\subfigure[\label{TDSCFigure3}]{\epsfig{figure=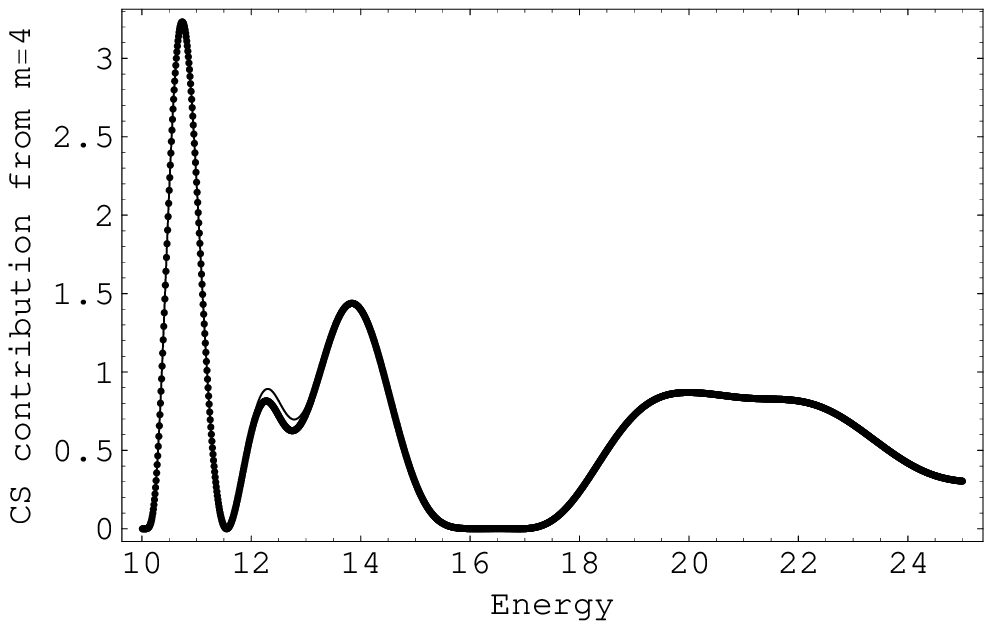,height=4.2cm}}
\subfigure[\label{TDSCFigure4}]{\epsfig{figure=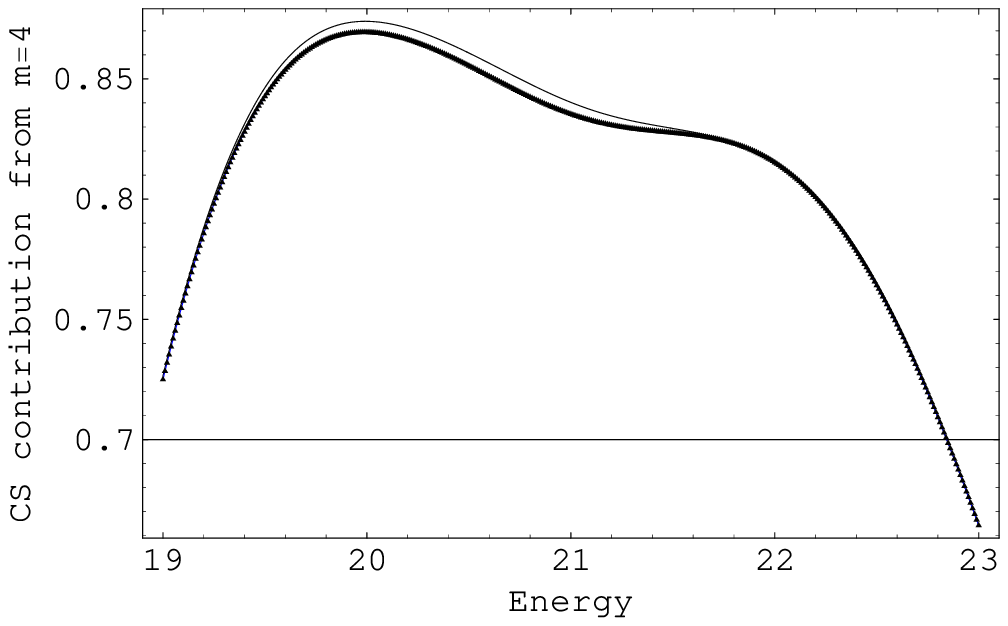,height=4.2cm}}
\end{center}
\caption[Total Scattering Cross-sections]{Total Scattering Cross-sections. Energy ranges from just above the well to 50 units higher. Well height was $V=10$ and the radius was $R=\sqrt{20}$. For Figure (a) $N=10$ and for Figure (b) $N=1000$. The solid line indicates the commutative cross-section, whereas the dots indicate the non-commutative values. For Figure (c) and (d) $N=1000$. In (c) and (d) we have plotted the contribution to the total cross-section for angular momentum, $m=4$, only. Figure (c) displays the cross-section over the first 25 units of energy and Figure (d) is a close up view for energies between 19 and 23. For (c) and (d) the solid line indicates the commutative case and the dots/triangles indicate the non-commutative case.}
\end{figure}

Due to the chosen values of $N$ in Figures \ref{TDSCFigure1} and \ref{TDSCFigure2}, we have that $\theta = \frac{20}{21}$ in Figure \ref{TDSCFigure1} and $\theta = \frac{20}{2001}$ in Figure \ref{TDSCFigure2}. As one can see, the non-commutative cross-section agrees much better with the commutative one as $\theta$ becomes smaller. This is of course what one would expect, since the non-commutative case must go over into the commutative one as $\theta\rightarrow 0$. Not shown in the graph is that the cross-section goes to infinity as the energy approaches the well height, i.e. $k \rightarrow 0$, in both the commutative and non-commutative case due to the $\frac{1}{k}$ appearing in the total cross-section. Figures \ref{TDSCFigure3} shows the contribution to the total cross-section from the $m=4$ angular momentum sector, where Figure \ref{TDSCFigure4} is a close up view of \ref{TDSCFigure3} near twenty units of energy. The deviations from the commutative case occur where the phase shifts differ as we saw in Figures \ref{PhaseShiftFigureSub2} and \ref{PhaseShiftFigureSub3}. In both graphs we see that there is a slightly lower contribution to the cross-section in the non-commutative case.

\section{Conclusion}
We have used the non-commutative formalism set out in \cite{Scholtz1} and the definition of the non-commutative well \cite{Scholtz2} to calculate bound state energies, phase shifts and the total cross-section of such a well. A more rigorous (than that used in \cite{Scholtz2}) derivation of the relation between the position and Fock space representations of $\hat{\psi}$ was also given for use when calculating matching conditions in the scattering region. In the numerical results for the bound states we observed a splitting in the energy of states with negative and positive angular momentum caused by non-commutativity. In the case of phase shifts and the total cross-section, we saw that the non-commutative results converge towards the commutative ones in the limit of $\theta \rightarrow 0$. However, from the data on phase shifts we saw that the convergence is not uniform for all energies. We note that mathematically the limit $\theta \rightarrow 0$ is similar to the quasi-classical limit $\hbar \rightarrow 0$, which is known not to be smooth in general. The physical origin of this is, however, not yet fully understood and bears more investigation.

\end{document}